\documentclass[a4paper,11pt,oneside,openany]{article}
\usepackage{graphicx}
\usepackage{color}
\usepackage{wrapfig}
\usepackage{comment}
\usepackage[top=20truemm,bottom=20truemm,left=20truemm,right=20truemm]{geometry}
\usepackage{amsmath}
\usepackage{authblk}
\usepackage{url}
\usepackage{color}
\usepackage[left]{lineno}

\title{First results on the search for the Galactic Center Excess in the sub-GeV band with the emulsion telescope in GRAINE 2023}

\author[1,2,$\dag$]{Yuya Nakamura}\author[4]{Shigeki Aoki}\author[4]{Takashi Azuma}\author[2]{Hirotaka Hayashi}\author[2]{Yudai Isayama}\author[5]{Atsushi Iyono}\author[4]{Takumi Kato}\author[2]{Tsuyoshi Kawahara}\author[6]{Kohichi Kodama}\author[2]{Ryosuke Komatani}\author[2]{Masahiro Komatsu}\author[2]{Masahiro Komiyama}\author[2]{Hideyuki Minami}\author[2]{Kunihiro Morishima}\author[5]{Fumiya Murakami}\author[2]{Shogo Nagahara}\author[2]{Naotaka Naganawa}\author[2]{Mitsuhiro Nakamura}\author[2]{Tomoaki Nakamura}\author[2]{Noboru Nakano}\author[2,3,1]{Toshiyuki Nakano}\author[7,8]{Kazuma Nakazawa}\author[4]{Miyuki Oda}\author[4]{Kazuhiro Okamoto}\author[3,1,2]{Hiroki Rokujo}\author[2]{Osamu Sato}\author[2]{Kai Shimizu}\author[2]{Amon Suganami}\author[5]{Yuki Sugi}\author[2]{Kou Sugimura}\author[4]{Satoru Takahashi}\author[2]{Ikuya Usuda}\author[2]{Saya Yamamoto}\author[4]{Jun Yamashita}\author[4]{Mayu Yamashita}\author[4]{Shoma Yoneno}\author[7,$\footnote{present address: RIKEN Nishina Center, RIKEN, 2-1 Hirosawa, Wako, Saitama 351-0198, Japan}$]{Masahiro Yoshimoto}

\affil[1]{Kobayashi-Maskawa Institute for the Origin of Particles and the Universe (KMI), Nagoya University, Furo-cho, Chikusa-ku, Nagoya, 464-8602, Japan}
\affil[2]{Graduate School of Science, Nagoya University, Furo-cho, Chikusa-ku, Nagoya, 464-8602, Japan}
\affil[3]{Institute of Materials and Systems for Sustainability (IMaSS), Nagoya University, Furo-cho, Chikusa-ku, Nagoya, 464-8601, Japan}
\affil[4]{Graduate School of Human Development and Environment, Kobe University, 3-11 Tsurukabuto, Nada-ku, Kobe, 657-8501, Japan}
\affil[5]{Graduate School of Science and Engineering, Okayama University of Science, 1-1 Ridaicho, Kita-ku, Okayama, 700-0005, Japan}
\affil[6]{Faculty of Education, Aichi University of Education, 1 Hirosawa, Igaya-cho, Kariya, 448-8542, Japan}
\affil[7]{Faculty of Education, Gifu University, 1-1 Yanagido, Gifu, 501-1193, Japan}
\affil[8]{The Research Institute of Nuclear Engineering (RINE), University of Fukui, 1-3-33 Kanawa, Tsuruga, Fukui, 914-0055, Japan}
\affil[$\dag$]{E-mail: ynakamura@flab.phys.nagoya-u.ac.jp}

\date{}

\begin{document}
\maketitle

\newpage

\begin{abstract}
{The Galactic Center Excess (GCE) is an unexplained excess of gamma-ray emission from the Galactic Center. While dark matter annihilation and millisecond pulsars have been proposed as candidate origins of the GCE, its origin remains unclear due to the large uncertainties associated with the Galactic diffuse gamma-ray emission in current observational data. The GRAINE experiment aims to reveal the origin of the GCE using an emulsion gamma-ray telescope with high angular resolutions of 1$^\circ$ at 100 MeV and 0.1$^\circ$ at 1 GeV. The high angular resolution enables observations of a region close to the Galactic Center, where the GCE intensity is high, with reduced contamination from diffuse gamma-ray emission. The GRAINE experiment observed the region around the Galactic Center for the first time during its 2023 balloon flight. In this study, we search for the GCE in a small region near the Galactic Center using the GRAINE 2023 flight data. In particular, rather than focusing on the spectral peak of the GCE at $\sim$2 GeV, we focused on the energy range below 300 MeV, where the spectral differences between the dark matter annihilation and millisecond pulsar scenarios are more pronounced. First, using the Galactic latitude distribution integrated over Galactic longitudes within $\pm$40$^\circ$, we detected gamma-ray emission along the Galactic plane for the first time with GRAINE, with a significance of 6.2$\sigma$. This confirms that the GRAINE observations of the region around the Galactic Center achieved the expected performance. We then searched for the GCE within 1$^\circ$ of the Galactic Center in the 75--300 MeV energy range. Although no significant excess was observed, we obtained an upper limit on the GCE flux of 1.70$\times$10$^{-7}$ GeV cm$^{-2}$ s$^{-1}$ at the 2$\sigma$ confidence level for the 1$^\circ$-radius ROI centered on the Galactic Center, based on a direct observation of the narrow region around the Galactic Center. This observation requires high angular resolution and represents a unique result from GRAINE. The obtained upper limit is consistent with the GCE flux near the Galactic Center, which was estimated from existing Fermi-LAT observations using a wide ROI and assuming an NFW profile. Although the current upper limit constrains some models, the available statistics are still insufficient to distinguish between the dark matter annihilation and millisecond pulsar scenarios, and both remain consistent with the current results. We also estimated the projected sensitivity of future GRAINE experiments based on the present observation and demonstrated their potential to probe the origin of the GCE by comparing the projected sensitivity with the predicted GCE spectra.}
\end{abstract}

\newpage
\tableofcontents
\newpage

\section{Introduction}
The Galactic Center is a particularly distinctive region of our Galaxy. It hosts a supermassive black hole, as well as dense interstellar matter and numerous astrophysical objects, making it an important target for observations across a broad range of astrophysical fields. Moreover, it is thought to contain a large amount of dark matter (DM), making the Galactic Center an important target for indirect dark matter searches. Over the past two decades, our understanding of the Galactic Center region has advanced through high-energy gamma-ray observations with the Fermi Large Area Telescope (Fermi-LAT)\cite{fermi}. These observations have revealed intriguing and unresolved phenomena, including a giant bubble structure thought to have been formed by past activity of the central black hole \cite{fermibubble} and an unidentified gamma-ray excess, known as the Galactic Center Excess (GCE)\cite{GCE1}\cite{GCE2}. 

The GCE is an unexpected gamma-ray emission discovered by the Fermi-LAT, characterized by an extended spatial morphology and a spectral peak at $\sim$1--3 GeV in $E^2 dN/dE$\cite{GCE1}\cite{GCE2}. Various analyses of Fermi-LAT data have strongly suggested that the excess is not an accidental feature arising from systematic uncertainties. More recently, the DArk Matter Particle Explorer (DAMPE) has also observed a similar excess, further strengthening the evidence for its existence\cite{DAMPE}. Two scenarios have been frequently discussed as possible origins of this excess. One is that the excess is produced by dark matter annihilation in the Galactic Center, where dark matter is expected to be abundant\cite{GCEDM1}\cite{GCEDM2}\cite{GCEDM3}\cite{GCEDM4}\cite{GCEDM5}. It has been reported that the observed spectrum can be explained by dark matter particles with masses of several tens of GeV annihilating through channels such as $\chi\chi\rightarrow b\overline{b}$, $c\overline{c}$, and $\tau^{+}\tau^{-}$, with a velocity-averaged annihilation cross section of $\sim$10$^{-26}$ cm$^3$ s$^{-1}$. This value is of particular interest because it is consistent with the thermally averaged annihilation cross section required to produce the observed abundance of thermal relic dark matter. In addition, the spatial extent of the excess has been reported to be consistent with the expected spatial distribution of dark matter in the Milky Way, described by an Navarro-Frenk-White (NFW) profile. The other scenario is that the GCE originates from a population of unresolved millisecond pulsars (MSPs)\cite{GCEMSP1}\cite{GCEMSP2}\cite{GCEMSP3}\cite{GCEMSP4}\cite{GCEMSP5}. This scenario is supported by the spectral shape of the GCE, which is consistent with the characteristic spectra of MSPs, and by studies suggesting that the spatial distribution of bulge MSPs inferred from the observed distribution of low-mass X-ray binaries (LMXBs) can be consistent with the spatial morphology of the GCE. The extended spatial distribution and spectrum of the gamma-ray excess can thus be explained by these two fundamentally different scenarios, and its origin remains unresolved. A major source of this uncertainty is the modeling of the Galactic diffuse gamma-ray emission.

In both the dark matter and MSP scenarios, the excess emission is expected to be strongly concentrated toward the Galactic Center. Although the diffuse gamma-ray emission is also intense near the Galactic Center, the GCE is expected to be more strongly concentrated toward the Galactic Center than the diffuse emission. Therefore, restricting the ROI to a small region around the Galactic Center can increase the GCE-to-diffuse emission ratio and reduce the associated model-dependent uncertainties. However, the limited angular resolution of the Fermi-LAT (5$^\circ$
 at 100 MeV and 0.8$^\circ$ at 1 GeV) limits how narrowly the ROI can be defined around the Galactic Center. Therefore, improving the angular resolution is crucial for reducing contamination from diffuse gamma-ray emission and determining the origin of the GCE.

The Gamma-Ray Astro-Imager with Nuclear Emulsion (GRAINE) project is a balloon-borne gamma-ray telescope experiment using nuclear emulsion films to observe cosmic gamma rays in the energy range of 10 MeV--100 GeV\cite{GRAINE}. Nuclear emulsion films are three-dimensional tracking detectors with sub-micron spatial resolution, enabling precise measurements of the electron and positron tracks produced by gamma-ray pair production. This provides GRAINE with an excellent angular resolution of 1$^\circ$ at 100 MeV and 0.1$^\circ$ at 1 GeV, as well as sensitivity to gamma-ray polarization\cite{GRAINE_polar}. In the 2018 balloon experiment (GRAINE 2018), GRAINE detected the Vela pulsar in the sub-GeV energy range with better angular resolution than the Fermi-LAT, demonstrating the performance of the telescope in the sub-GeV energy range\cite{GRAINE 2018_1}\cite{GRAINE 2018_2}. In the 2023 balloon flight (GRAINE 2023), GRAINE observed the region around the Galactic Center for the first time. In this study, we report the results of a search for the GCE in a region confined to the vicinity of the Galactic Center using the GRAINE 2023 data. In particular, rather than focusing on the $\sim$2 GeV energy range, where the spectral energy distribution (SED), expressed as ($E^2 dN/dE$), of the GCE is expected to peak, we focus on the energy range below 300 MeV. In the dark matter annihilation scenario, the predicted spectrum peaks at an energy related to the dark matter mass and decreases rapidly below $\sim$2 GeV. In contrast, the millisecond pulsar scenario predicts a power-law-like spectral shape, resulting in a relatively hard spectrum even below $\sim$2 GeV. Therefore, the GCE flux in the few-hundred-MeV energy range is particularly important for distinguishing between these two scenarios. However, in current Fermi-LAT observations, the angular resolution is substantially poorer in this energy range than at GeV energies, making it difficult to study the GCE at several hundred MeV, where the systematic uncertainties are particularly large. In contrast, GRAINE achieves an angular resolution of $\sim$0.5$^\circ$ at 200 MeV, comparable to that of the Fermi-LAT at 2 GeV, substantially reducing contamination from diffuse gamma-ray emission caused by limited angular resolution. Taking advantage of this high angular resolution, we searched for the GCE in the 75--300 MeV energy range using the GRAINE 2023 flight data, restricting the region of interest (ROI) to within 1$^\circ$ of the Galactic Center. This represents a unique observation by GRAINE, as a direct search for the GCE in such a small region around the Galactic Center is difficult with the Fermi-LAT, whose angular resolution at 200 MeV is $\sim$3$^\circ$. Section 2 provides an overview of GRAINE 2023. Section 3 describes the data analysis and the results for the Vela pulsar observed in the same experiment, which are used to verify the basic performance of the telescope in GRAINE 2023. Section 4 presents the analysis of gamma-ray emission from the Galactic plane as a performance verification of the data obtained around the Galactic Center, which was observed by GRAINE for the first time. Finally, Section 5 presents the results on the search for the GCE.

\section{GRAINE 2023 experiment}
\begin{figure}
\center
\includegraphics[bb=0 0 970 831, width=.6\textwidth]{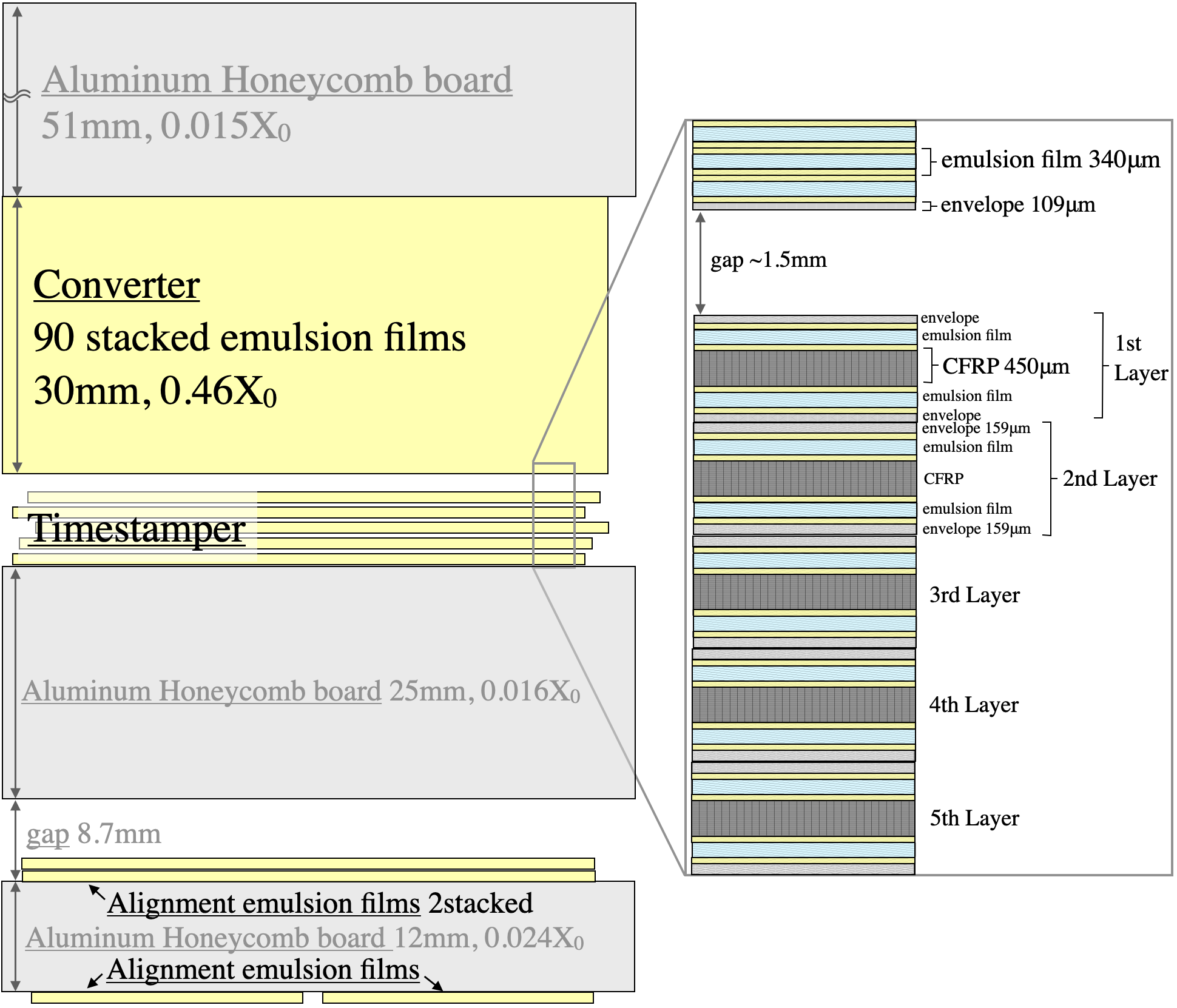}
\caption{Schematic of the emulsion chamber configuration used in GRAINE 2023.}
\label{fig:detector}
\center
\end{figure}

We describe the detector configuration of GRAINE 2023. The GRAINE emulsion gamma-ray telescope consists of an emulsion chamber and an attitude monitor. Figure \ref{fig:detector} shows the configuration of the emulsion chamber used in GRAINE 2023. The emulsion chamber consists of converters, which detect gamma rays and measure their directions and energies, and time stampers, which provide timing information for the detected tracks. Two newly developed time stamper units were installed to increase the detector size and improve its time resolution while reducing its weight\cite{GRAINE_shifter1}\cite{GRAINE_shifter2}. The emulsion films used in the experiment are 25 cm $\times$ 50 cm in size, consisting of a 200 $\mu$m-thick plastic base film with 70 $\mu$m-thick emulsion layers applied to both sides. Each converter consists of a stack of 90 emulsion films. Two converters are attached to each aluminum honeycomb board, and five honeycomb boards, corresponding to 10 converters, are installed for each time stamper unit. In addition, alignment films are placed downstream of the time stampers and vacuum-packed together with an aluminum honeycomb board to provide a reference for calibrating the planarity of the converter films. The total aperture area is 2.5 m$^2$, and the emulsion chamber is installed in a pressure-vessel gondola (see \cite{GRAINE_gondola} for details). The aperture area is approximately 6.6 times as large as that of the GRAINE 2018 (0.38 m$^2$). The attitude monitor, mounted outside the pressure vessel, consists of CMOS cameras (Triton TRI028S-MC) and the same lenses, low-pass filters, and baffle hoods used in GRAINE 2018 \cite{GRAINE 2018_1}. Three units are installed, each oriented in a different azimuthal direction, with 90$^\circ$ separation between adjacent units. The cameras continuously acquire images of stars in visible light, and the telescope attitude is determined after the flight by matching the observed star images with a star catalog. The incident direction and energy of each gamma ray are reconstructed by measuring the angles of the electron and positron tracks near the gamma-ray conversion point and the angular deflections caused by multiple Coulomb scattering as the tracks traverse multiple emulsion films. By associating each event with its detection time and the corresponding telescope attitude, the reconstructed gamma-ray directions are converted into celestial coordinates.

\begin{figure}
\center
\includegraphics[bb=0 0 1429 520, width=.8\textwidth]{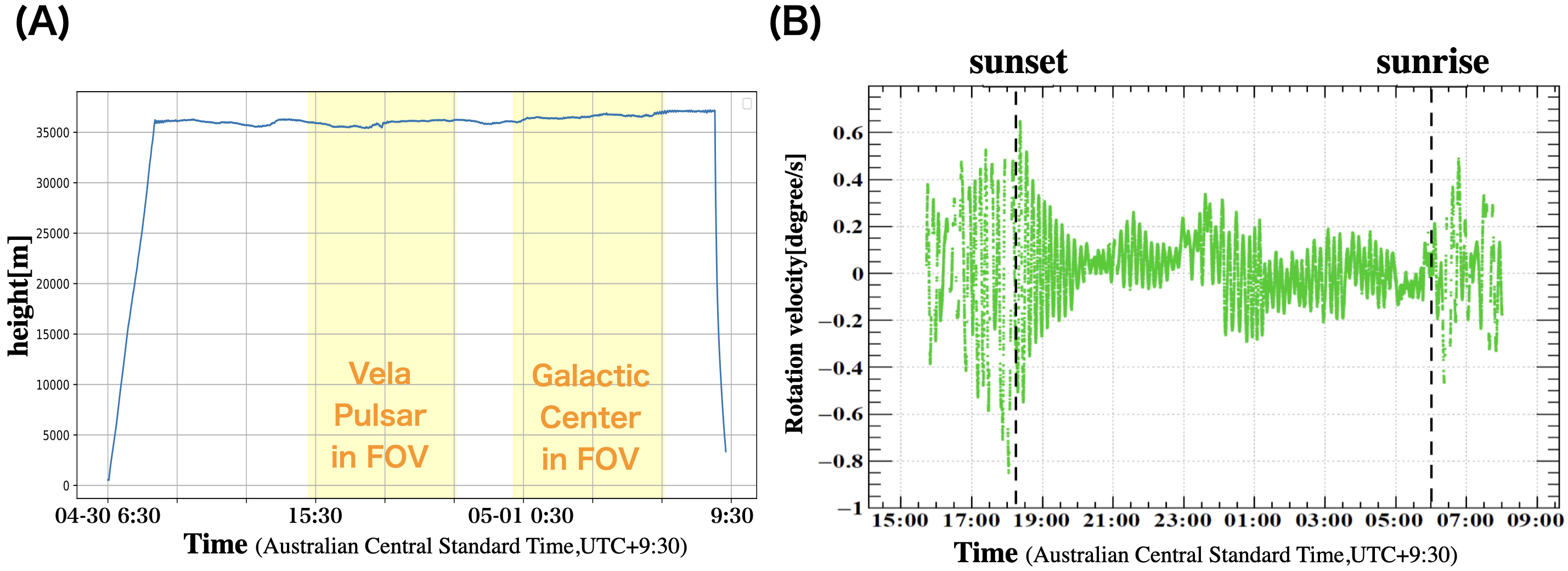}
\caption{(A) Altitude profile during the observation. (B) Azimuthal rotation rate of the telescope.}
\label{fig:height_rot}
\center
\end{figure}

Details of the GRAINE 2023 flight are described in \cite{GRAINE_gondola}. In brief, the balloon was launched at 06:32 on April 30, 2023 (Australian Central Standard Time, UTC+9:30), entered the level-flight phase at approximately 08:30, and conducted observations for 24.3 h at an altitude of approximately 36 km. As representative observation targets, the Vela pulsar was within the telescope's field of view (zenith angle $<$45$^\circ$) from 15:11:17 to 21:34:32 on April 30, while the Galactic Center was within the field of view from 00:04:10 to 06:35:24 on May 1. Figure \ref{fig:height_rot} shows (A) the altitude profile during the observation and (B) the rotational velocity of the telescope in azimuth, derived from the attitude-monitor data.

\section{Data analysis and basic performance evaluation}
The basic analysis procedure is the same as that described in \cite{GRAINE 2018_1}. Here, we focus on the dataset used in the present analysis, the performance of the newly developed time stamper system, and the basic performance evaluation based on the analysis of the Vela pulsar.
\subsection{Data set}\label{sec:ds}
Approximately 6 hours after the balloon entered the level-flight phase, the operation mode of the time stampers was changed to the high-time-resolution mode for astronomical observations. We used approximately 16 hours of data taken after the mode change (April 30, 2023, 15:57:29 to May 1, 07:44:07) and selected approximately 1.5$\times$10$^6$ gamma-ray events with incident zenith angles within 45$^\circ$ and energies above 75 MeV. Gamma rays produced secondarily within the detector were identified and removed based on the time coincidence between charged secondary particles and gamma rays. For the identification of electron-positron pairs, the 90 stacked emulsion films are searched using groups of eight consecutive films, which are shifted one film at a time through the stack. The upstream three films serve as veto layers, while the downstream five films are used to search for two closely spaced tracks produced at an intermediate point in the stack. In the present analysis, we also applied a new processing method that extends the number of films used for track searching from 5 to 25. This method improves the detection efficiency for high-energy events with a small opening angle between the electron and positron tracks, for which the two tracks cannot be separately detected when only a small number of films are used for the search. This modification provides no significant improvement in the detector performance below 300 MeV, which is the primary energy range of this study. Details of the gamma-ray detection method are described in \cite{GRAINE_gamma}.

\begin{figure}
\center
\includegraphics[bb=0 0 1180 830, width=.7\textwidth]{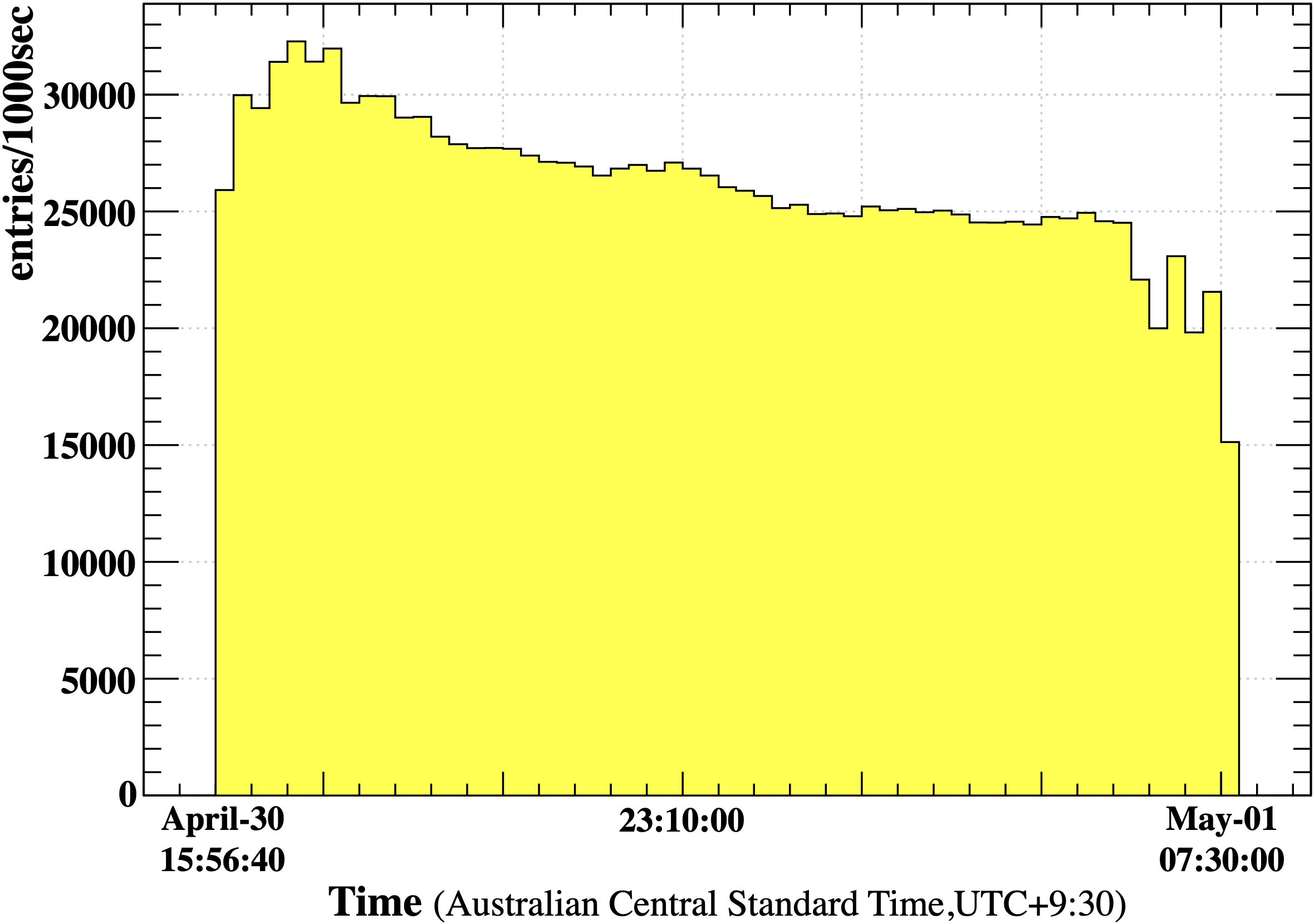}
\caption{Count rate of selected gamma-ray events.}
\label{fig:gtime}
\center
\end{figure}

Figure \ref{fig:gtime} shows the count rate of the selected gamma-ray events. The slight decrease in the number of detected events with time is considered to be caused by a decrease in the sensitivity of the emulsion films, particularly those used in the time stamper, in the low-temperature environment, resulting in a decrease in the efficiency of assigning timing information to the events. A decrease in sensitivity was also observed for the films used in the converter; however, their sensitivity remained sufficiently high for tracks to be recognized by the emulsion film scanning system, and therefore the observation performance was not affected. In future experiments, this issue is expected to be addressed by increasing the initial sensitivity of the emulsion films used in the time stamper.

\begin{figure}
\center
\includegraphics[bb=0 0 1434 605, width=.8\textwidth]{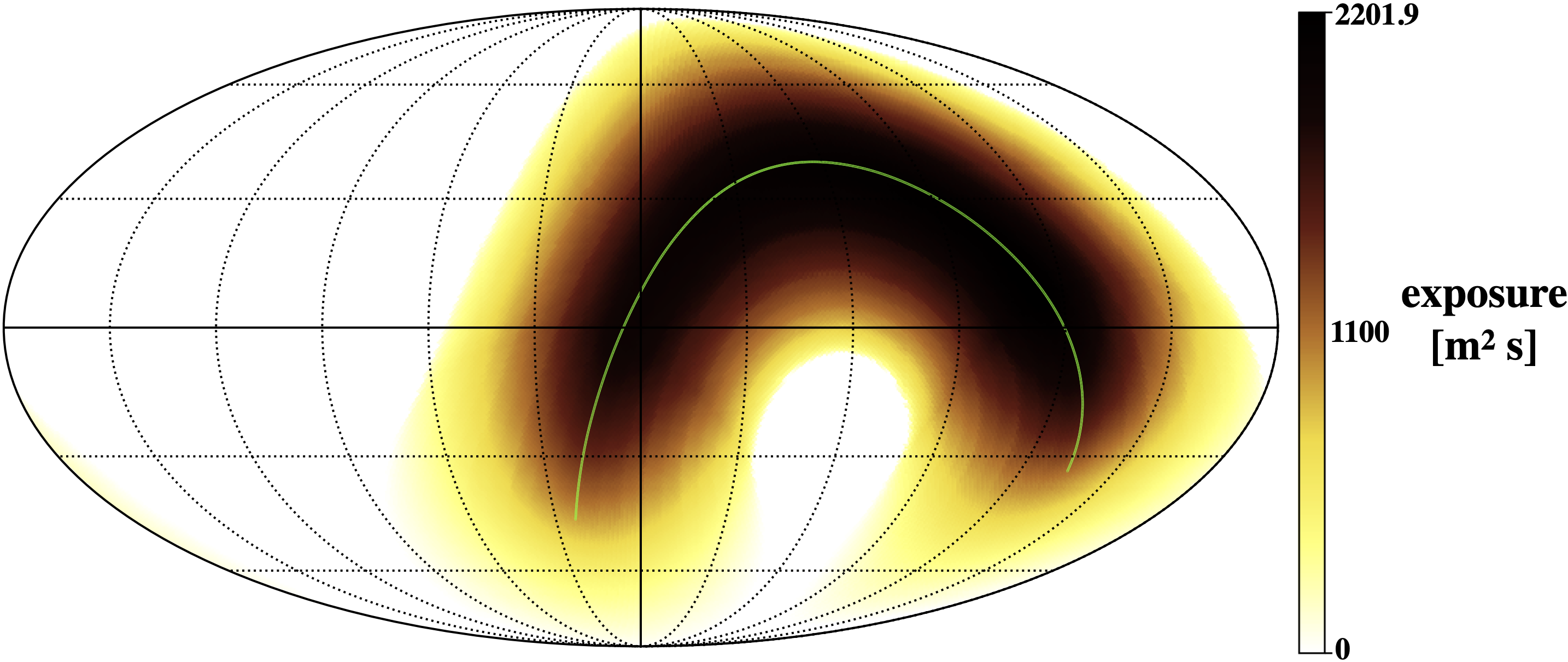}
\caption{Exposure map (effective area $\times$ observation duration) at 200 MeV. The green line shows the Galactic coordinates of the telescope zenith at each time.}
\label{fig:exposure}
\center
\end{figure}

The various efficiencies and detector response were evaluated following the same procedure as in the previous analysis \cite{GRAINE 2018_1}. The selection efficiency for electron and positron pair production events was evaluated using simulations with Geant4 \cite{GRAINE_gamma}\cite{geant}. The timing-assignment efficiency for the selected events was evaluated using the flight data by checking whether, when one track of an electron-positron pair was assigned a timing, the other track could also be assigned a timing consistent with it within the timing resolution. The timing resolution is evaluated in the next subsection. The fraction of the observation time for which the telescope attitude was successfully determined was greater than 98\% over the entire observation period and approximately 100\% during nighttime. Figure \ref{fig:exposure} shows the exposure map at 200 MeV, calculated from the evaluated detector response, where the exposure is defined as the effective area multiplied by the observation duration.

Most of the selected gamma-ray events are atmospheric gamma rays produced by interactions between cosmic rays and the atmosphere, mainly through the decay of neutral pions. We therefore constructed a background dataset (BG) for subtracting this component in the analysis. In constructing the BG, random background events were generated to reproduce the time, azimuthal, and incident-angle distributions of the selected gamma-ray data, taking into account the time dependence due to changes in the observation altitude, the azimuthal dependence due to the east-west effect of cosmic rays, and the incident-angle dependence. The BG was normalized by comparison with the flight data after masking the region within $\pm$1.5$^\circ$ in Galactic latitude and the region within $\pm$2$^\circ$ in both Galactic longitude and latitude around the position of the Vela pulsar. The normalization uses gamma-ray data over the entire field of view outside these masked regions, where atmospheric gamma rays are dominant. Therefore, any bias in the normalization factor due to the contribution from cosmic gamma rays is negligible.

\subsection{Performance of the newly developed time stamper}
\begin{figure}
\center
\includegraphics[bb=0 0 1097 833, width=.6\textwidth]{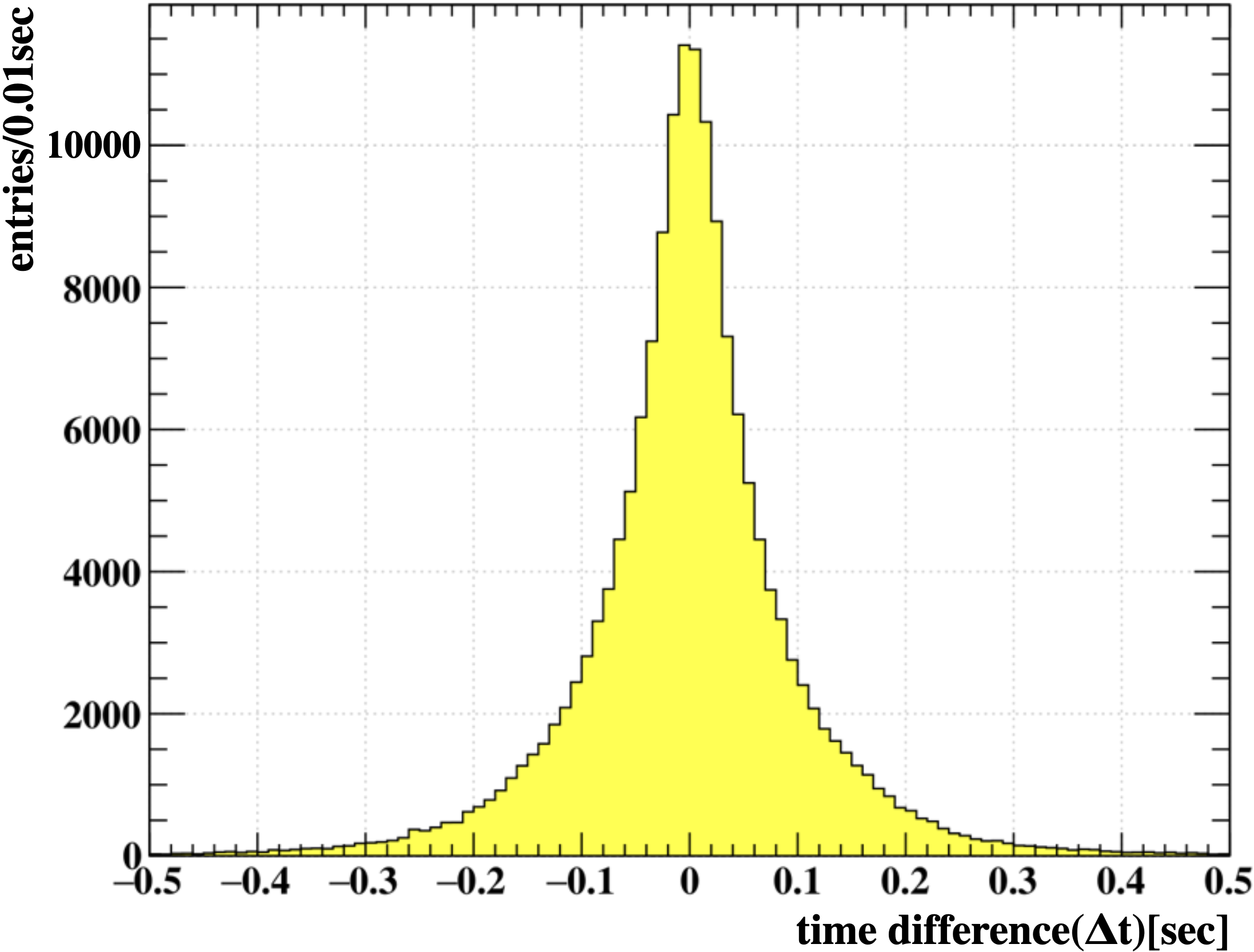}
\caption{The distribution of the time difference, $\Delta t$, between the timings independently assigned to the two tracks of each electron-positron pair.}
\label{fig:timeres}
\center
\end{figure}

Details of the configuration of the newly developed time stamper introduced for GRAINE 2023 are described in \cite{GRAINE_shifter2}. Here, we describe the evaluation of its time resolution using the flight data. Figure \ref{fig:timeres} shows the distribution of the time difference, $\Delta t$, between the timings independently assigned to the two tracks of each electron-positron pair, using a subset of the dataset described above. We define $\sigma_{\Delta t}$ as the width centered at $\Delta t$=0 that contains 68\% of the events in this distribution. The timing determination precision for a single track is then $\sigma_{\Delta t}/\sqrt{2}$=0.054 s. When timing information could be assigned to both tracks of an electron-positron pair, the timing assigned to the track with the higher momentum was adopted as the event arrival time. We estimated the resulting uncertainty in the reconstructed arrival direction for astronomical observations due to the timing determination precision. As shown in Figure \ref{fig:height_rot}(B), the rotational speed of the telescope was less than 0.51$^\circ$ s$^{-1}$ for 99\% of the entire observation period. Since this is the rotational speed in the azimuthal direction, the effect is smaller for incident directions closer to the zenith. Thus, even at the edge of the telescope field of view, corresponding to a zenith angle of 45$^\circ$, the maximum effective rotational speed is 0.51$\times\cos(45^\circ)$=0.36$^\circ$ s$^{-1}$. Using these values, the maximum change in the reconstructed gamma-ray direction due to the timing uncertainty is 0.054$\times$0.36=0.020$^\circ$. This is sufficiently smaller than the angular resolution of the emulsion films, 0.1$^\circ$ at 1 GeV, in the primary observation energy range of GRAINE 2023 below 1 GeV, and can therefore be safely neglected.

\subsection{Vela pulsar}\label{sec:vela}
\begin{figure}
\center
\includegraphics[bb=0 0 1454 725, width=.8\textwidth]{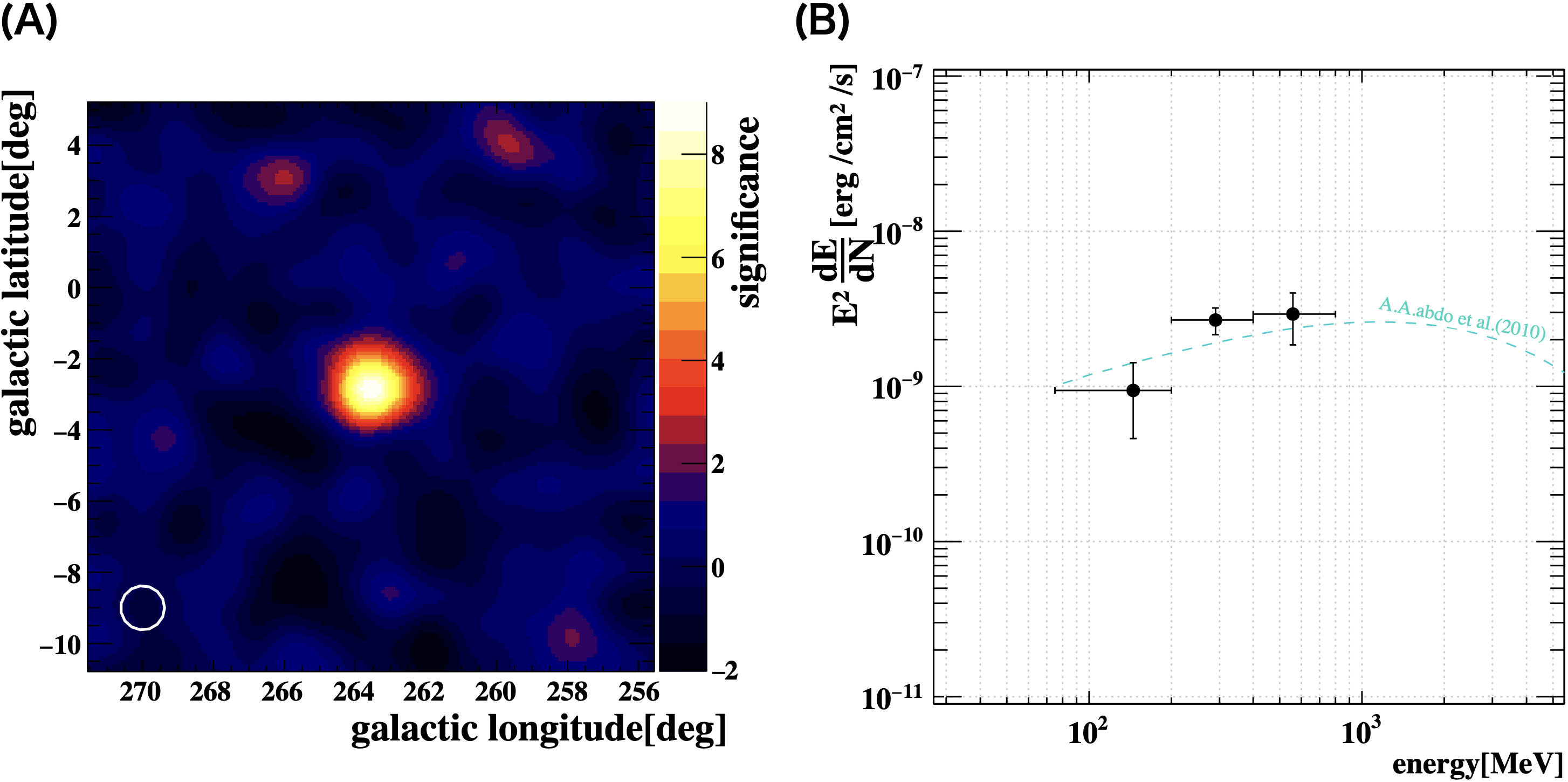}
\caption{Results of the Vela pulsar analysis. (A) Significance map in Galactic coordinates. The white circle indicates the 68\% containment radius of the signal. (B) Spectral energy distribution (SED). The light-blue dashed line shows the best-fit SED obtained from Fermi-LAT observations \cite{fermiVela}.}
\label{fig:vela}
\center
\end{figure}

Prior to the analysis of the region around the Galactic Center, we evaluated the performance of the GRAINE 2023 flight data by analyzing the Vela pulsar, which has been successfully observed in previous GRAINE flights. The basic analysis procedure is the same as that used for the Vela pulsar analysis in GRAINE 2018 \cite{GRAINE 2018_1}. Figure \ref{fig:vela}(A) shows the significance map in Galactic coordinates. After subtracting the BG, the data were smoothed with a Gaussian kernel with $\sigma$=0.6$^\circ$. The BG was smoothed in the same way, and the significance was calculated for each bin as signal/$\sqrt{BG}$. The energy range is defined as above 75 MeV without an upper bound. For some high-energy events, the multiple Coulomb scattering of the electron and positron tracks is too small to determine their momenta precisely, and only lower limits on their energies can therefore be assigned. After BG subtraction, we defined the total number of signal events as the number of events within 4$^\circ$ of the Vela pulsar. The radius containing 68\% of the signal events, together with its statistical uncertainty, was found to be 0.62$^{+0.07}_{-0.06}$ deg. A detailed evaluation of the angular resolution of the converter was presented in \cite{GRAINE 2018_2}, from which an angular resolution of approximately 0.57$^\circ$ is expected. This confirms that the GRAINE 2023 flight data achieved the expected angular resolution. The imaging performance obtained here is limited by the accuracy of the high-speed scanning system for the nuclear emulsion films. The imaging performance is expected to be improved to the design value of GRAINE, approximately 0.27$^\circ$, through reanalysis using a high-precision scanning system that targets only gamma-ray events (see \cite{GRAINE_precise} for details of the high-precision scanning system). Figure \ref{fig:vela}(B) shows the measured SED, which is consistent with previous observations by Fermi-LAT \cite{fermiVela}. The energy distribution is concentrated in the sub-GeV range: approximately 66\% of the detected events have energies below 500 MeV and approximately 90\% below 800 MeV. Thus, the analysis of the Vela pulsar confirms that the imaging performance and detector response of GRAINE in the sub-GeV energy range are consistent with the expected performance.

\section{Emission from the Galactic plane}\label{sec:gpe}
\begin{figure}
\center
\includegraphics[bb=0 0 797 792, width=.6\textwidth]{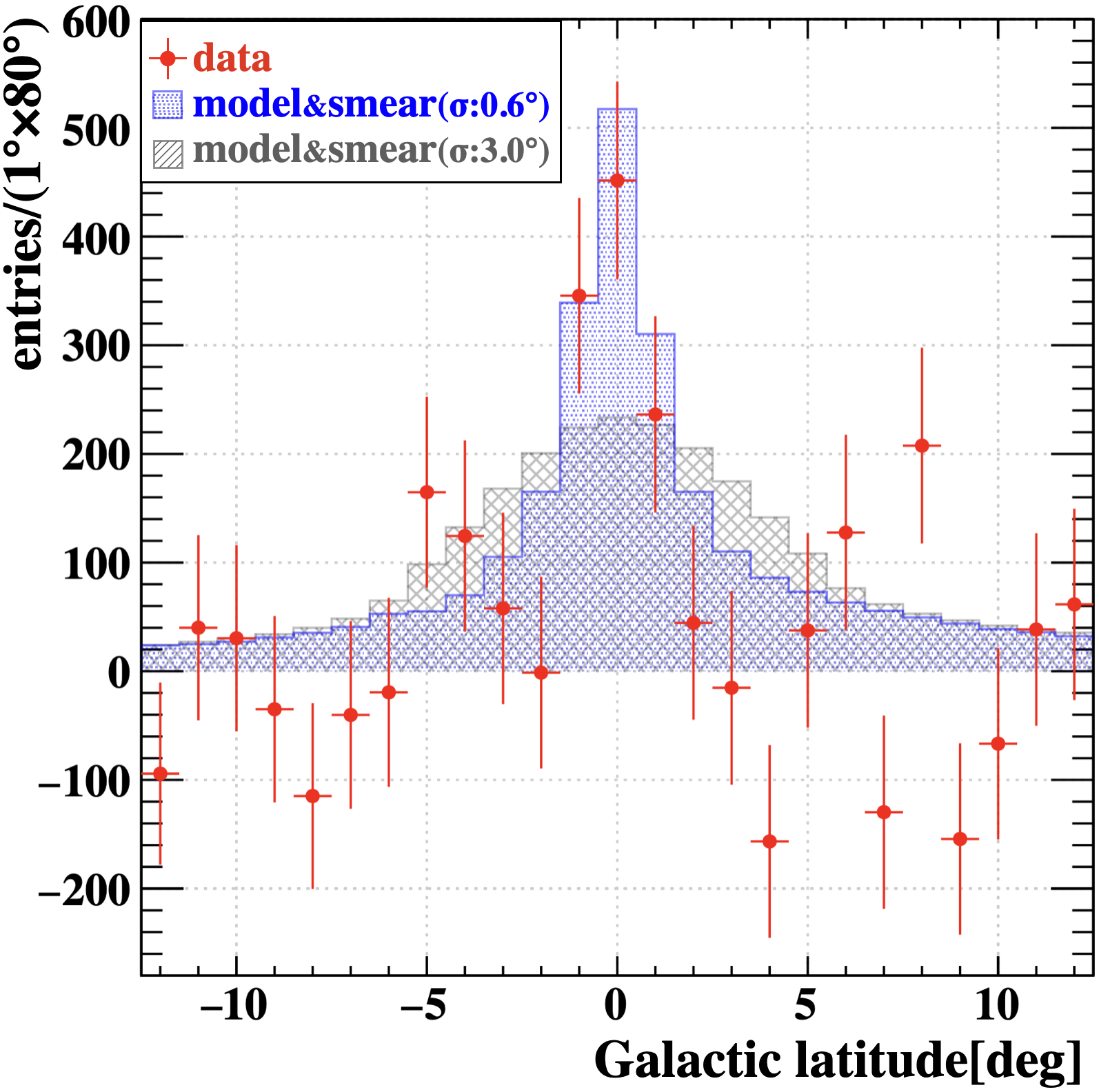}
\caption{Galactic latitude projection distribution within $\lvert l \rvert <$40$^\circ$ and $\lvert b \rvert <$12.5$^\circ$. The red points show the observed data above 75 MeV. Statistical uncertainties are assigned to the data points, and these uncertainties are dominated by the statistics of the events before BG subtraction. The blue histogram shows the expected distribution calculated from the publicly available Fermi-LAT models for the energy range of 75 MeV--10 GeV and smeared with a Gaussian of $\sigma$=0.6$^\circ$. The gray histogram shows the expected distribution smeared with a Gaussian of $\sigma$=3.0$^\circ$.}
\label{fig:diffuse}
\center
\end{figure}

We searched for gamma-ray emission along the Galactic plane using the data from the region around the Galactic Center. After subtracting the BG, we obtained the gamma-ray count distribution projected onto Galactic latitude, using events within $\pm$40$^\circ$ in Galactic longitude and $\pm$12.5$^\circ$ in Galactic latitude. The comparison between the observed distribution and the model prediction is shown in Figure \ref{fig:diffuse}. The uncertainties in the data are dominated by statistical uncertainties associated with the number of events before BG subtraction. For the model prediction, we first calculated the gamma-ray flux as a function of Galactic longitude and latitude in the energy range of 75 MeV--10 GeV using the publicly available Fermi-LAT diffuse gamma-ray model (gll\_iem\_v07), point-source catalog (gll\_psc\_v35), and isotropic gamma-ray model (iso\_P8R3\_SOURCE\_V3\_v1). The flux was then converted into the expected gamma-ray counts by applying the exposure map in each energy bin. The detector response used to calculate the exposure map was based on that validated by the spectral analysis of the Vela pulsar described in Section \ref{sec:vela}. The slight decrease in timing-assignment efficiency during the latter part of the observation, corresponding to the observation period around the Galactic Center, was also taken into account as suggested by the detected event rate in Figure \ref{fig:gtime}. Although the data include events above 75 MeV, the detection efficiency above 10 GeV is negligibly small with the current analysis method. Therefore, the model prediction was calculated for the energy range of 75 MeV--10 GeV. The slight mismatch between the energy binning of the publicly available models and that used in our analysis was corrected by linear interpolation. Finally, to account for the image broadening of events above 75 MeV, the positions of the expected gamma rays were smeared in Galactic longitude and latitude according to a Gaussian distribution with $\sigma\sim$0.6$^\circ$, based on the angular resolution obtained from the Vela pulsar analysis in the previous section. The expected gamma-ray counts were then obtained by projecting the smeared model onto Galactic latitude over the same region as that used for the data. For reference, we also show the result for a model smeared with $\sigma$=3.0$^\circ$, which is comparable to the Fermi-LAT angular resolution at 200 MeV.

The observed data were found to be consistent with the expected Galactic latitude distribution in terms of the total number of detected gamma rays, image broadening, and the peak at the Galactic plane (b=0$^\circ$). We calculated the detection significance of gamma-ray emission from the Galactic plane using a likelihood-ratio test based on Wilks' theorem \cite{wilks}. The expected Galactic latitude distribution was used as a template, and the test statistic (TS) was calculated from the likelihood of the best-fit model, $L$, and that of the null hypothesis, $L_0$, as TS=-2(ln$L_0$-ln$L$). The detection significance was then evaluated from the TS as follows\cite{TS}.
\begin{equation}
\sqrt{\rm{TS}}=\frac{\Sigma d_i s_i / \sigma_i^2}{\sqrt{\Sigma s_i^2 /  \sigma_i^2}}
\end{equation}
Here, $d_i, s_i, \sigma_i$ denote the data value, the fraction of the total template, and the uncertainty, respectively, in the $i$-th bin. Applying this method to the GRAINE 2023 data, we obtained a detection significance of 6.2$\sigma$. This result demonstrates, for the first time with GRAINE, a statistically significant detection of cosmic gamma-ray emission along the Galactic plane and confirms that the data obtained during the period when the Galactic Center was within the field of view exhibit the expected detector performance.

\section{Galactic Center Excess}
\begin{figure}
\center
\includegraphics[bb=0 0 925 813, width=.8\textwidth]{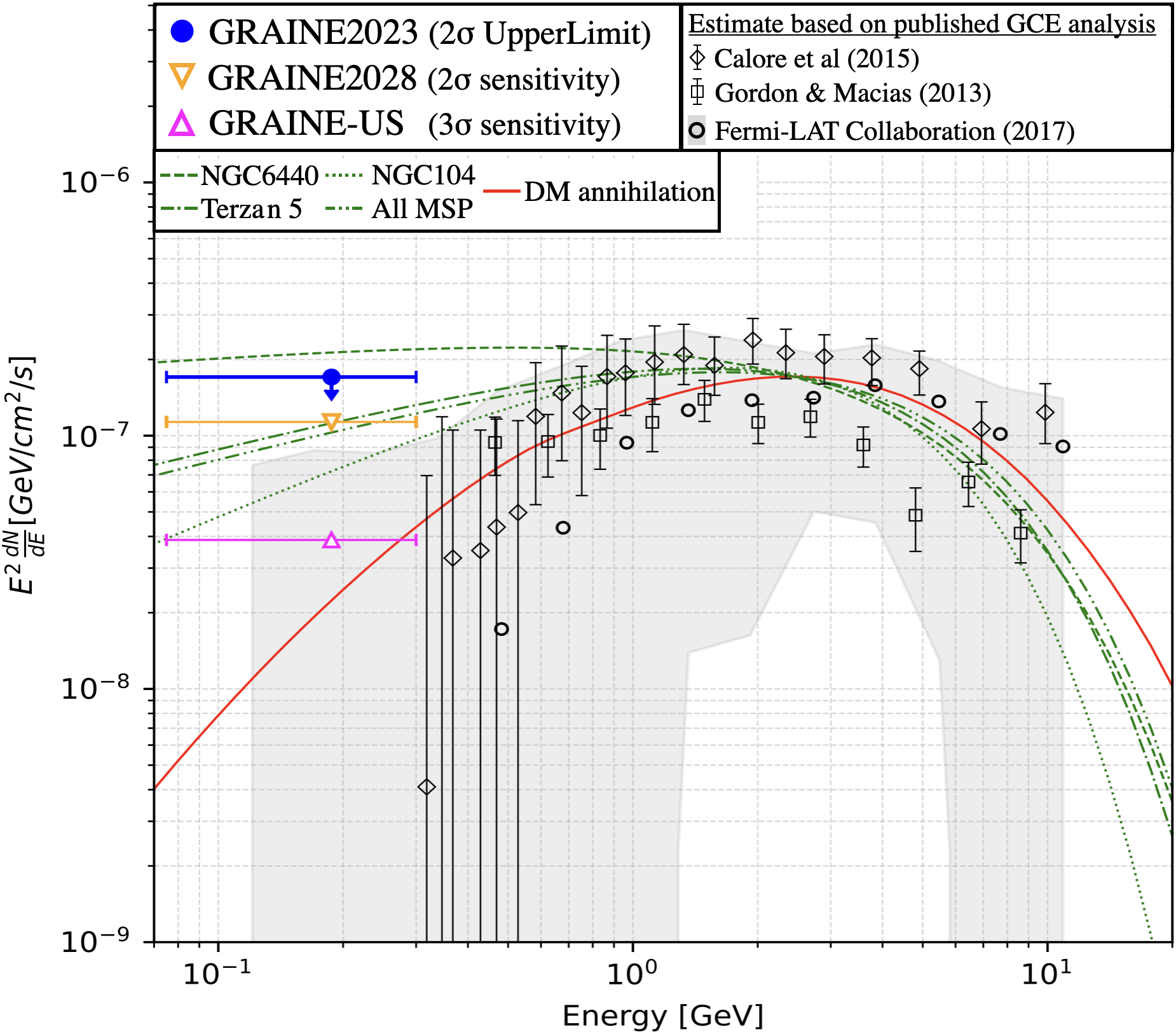}
\caption{The red solid line shows the spectrum assumed under the dark matter scenario (see the main text for details of the model parameters), while the green line shows the spectrum assumed under the millisecond-pulsar scenario, normalized to the integrated spectrum predicted under the dark matter scenario over 1.5--6 GeV. The green dashed, dotted, dash-dotted, and dash-double-dotted lines show the spectra for NGC 6440, NGC 104, Terzan 5, and the average spectrum of millisecond pulsars observed by Fermi-LAT, respectively. The black plot, error bars, and error band show the spectrum within 1$^\circ$ of the Galactic Center estimated by assuming a spatial distribution based on the results of a GCE analysis using a wide ROI. The open diamonds with error bars, open squares with error bars, and open circles with a gray error band show the results from Calore et al. (2015), Gordon and Macias (2013), and the Fermi-LAT Collaboration (2017), respectively. The blue filled circles with downward arrows show a conservative upper limit of 1.70$\times$10$^{-7}$ GeV cm$^{-2}$ s$^{-1}$ at the 2$\sigma$ confidence level for the 1$^\circ$-radius ROI centered on the Galactic Center, obtained from the GRAINE 2023 flight data. The horizontal error bars indicate the energy range used in the analysis. The open orange downward triangles show the projected sensitivity (2$\sigma$) for the next GRAINE balloon experiment (GRAINE 2028), assuming the same energy range as that used for the GRAINE 2023 analysis. The open magenta triangles show the projected sensitivity (3$\sigma$) for a future GRAINE experiment with an aperture area of 10 m$^2$ and an observation duration of one week.}
\label{fig:GCE}
\center
\end{figure}

In this section, we discuss the search for the Galactic Center Excess (GCE) within 1$^\circ$ of the Galactic Center in the energy range of 75--300 MeV. Figure \ref{fig:GCE} shows the spectra predicted under the dark matter and millisecond pulsar scenarios, the spectrum within 1$^\circ$ of the Galactic Center estimated based on the GCE spectrum obtained from an analysis of Fermi-LAT data using a wide region of interest (ROI), the GRAINE 2023 observation results, and the projected sensitivity of future GRAINE experiments. Details of each are described below.

\subsection{Spectral models for the GCE}
First, we calculated the predicted gamma-ray flux from the dark matter scenario for the GCE. The gamma-ray flux from dark matter annihilation is given by the following equation.
\begin{equation}\label{eq:NFW}
\frac{d\Phi}{dE}=\frac{1}{4\pi}\frac{\langle\sigma_{\rm{A}} v\rangle}{2m_{\rm{DM}}^2}\frac{dN}{dE}\int_\Omega\int_{l.o.s}\rho_{\rm{DM}}^2(r)drd\Omega
\end{equation}
Here, $m_{\rm{DM}}$ is the dark matter mass, $\langle\sigma_{\rm{A}} v\rangle$ is the velocity-averaged annihilation cross section, and $dN/dE$ is the differential gamma-ray yield per annihilation, which depends on the annihilation channel. $\rho_{\rm{DM}}$ is the dark matter density distribution, and the integration is performed along the line of sight and over the solid angle. For the dark matter density distribution, we adopted the NFW profile, which has been discussed as a model capable of explaining the GCE observed by Fermi-LAT. The NFW profile is expressed as follows.
\begin{equation}
\rho_{\rm{DM}}(r)=\rho_0\frac{(r/r_{\rm{s}})^{-r}}{(1+r/r_{\rm{s}})^{3-r}}
\end{equation}
Here, $\rho_0$ and $r_{\rm{s}}$ are the scale density and scale radius, respectively. Based on the dark matter density in the vicinity of the Solar System and the size of the Milky Way, typical values of 0.4 GeV cm$^{-3}$ and 20 kpc are adopted for $\rho_0$ and $r_{\rm{s}}$, respectively. $\gamma$ is the density parameter, and $\gamma$=1.26 has been discussed as a value that can explain the GCE observed by Fermi-LAT \cite{GCE_calore}; we adopt this value in this analysis. We evaluated the integral within an angular distance of 1$^\circ$ from the Galactic Center, assuming a distance of 8.5 kpc between the Galactic Center and the Solar System. For $dN/dE$, we used the $\chi\chi\rightarrow b\overline{b}$ channel from PPPC4DMID, which is based on simulations using PYTHIA, and adopted a dark matter mass of $m_{\rm{DM}}$=50 GeV \cite{PPPC4}. For $\langle\sigma_{\rm{A}} v\rangle$, we adopted 1.76$\times$10$^{-26}$ based on the analysis of Fermi-LAT data presented in \cite{GCE_fermi1}. Using these values and Equation \ref{eq:NFW}, we derived the spectrum within 1$^\circ$ of the Galactic Center, which is shown in Figure \ref{fig:GCE}.

The spectrum of millisecond pulsars can be inferred from the spectra of globular clusters. Since globular clusters are old and dense stellar systems, they contain abundant millisecond pulsars, and high-energy gamma-ray emission around the GeV energy range is thought to be predominantly produced by millisecond pulsars. Therefore, their spectra are often used as a reference when discussing the millisecond-pulsar origin scenario for the GCE. We determined the spectral shape based on observations of globular clusters by Fermi-LAT \cite{GClist}. The absolute normalization depends on the number of millisecond pulsars within the observation region (here, within 1$^\circ$ of the Galactic Center). For simplicity, we therefore normalized the spectrum to the integrated gamma-ray spectrum from the dark matter scenario over 1.5--6 GeV, as derived from the NFW profile described above. The 1.5--6 GeV range is a region where the uncertainty in the Fermi-LAT observations is relatively small. In addition, numerous GCE analyses have been performed assuming either the dark matter or millisecond-pulsar origin, and neither scenario has been ruled out. Figure \ref{fig:GCE} shows the predicted spectra for the millisecond-pulsar scenario, each based on a different spectrum: the average spectrum of Fermi-LAT-detected millisecond pulsars presented in \cite{AllMSP}, and the spectra of Terzan 5, NGC 6440, and NGC 104 presented in \cite{GClist}.

Based on the GCE spectra obtained from analyses of Fermi-LAT data using wide ROIs, we estimated the spectra within 1$^\circ$ of the Galactic Center, which is the region considered in this paper. We refer here to the results of Calore et al. (2015) \cite{GCE_fermi1}, Gordon and Macias (2013) \cite{GCE_fermi2}, and the Fermi-LAT Collaboration (2017) \cite{GCE_fermi3}. The ROIs used in these studies were 2$^\circ$ $< \lvert b \rvert <$ 20$^\circ$ and $\lvert l \rvert$ $<$ 20$^\circ$, a 7$^\circ$$\times$7$^\circ$ square centered on the Galactic Center, and a 10$^\circ$-radius region with the inner 2$^\circ$ effectively excluded by masking bright sources, respectively. Although the ROIs and analysis methods differ among these studies, we assumed an NFW spatial distribution with $\gamma$=1.26, calculated the $J$ factor for each ROI, and rescaled the spectra using the $J$ factor within 1$^\circ$ of the Galactic Center for normalization. The $J$ factor is given by the spatial integral in Equation \ref{eq:NFW}:
\begin{equation}
J=\int_\Omega\int_{l.o.s}\rho_{\rm{DM}}^2(r)drd\Omega
\end{equation}
The upper and lower bounds of the uncertainties reported in each study were also rescaled by the same factors. Figure \ref{fig:GCE} shows the estimated spectra after rescaling to the region within 1$^\circ$ of the Galactic Center.

\subsection{Results in GRAINE 2023}
We searched for the GCE in the GRAINE 2023 flight data by comparing the number of gamma-ray events detected within 1$^\circ$ of the Galactic Center with the expected number of events. The expected number of events consists of (1) the background (BG), dominated by atmospheric gamma rays, described in Section \ref{sec:ds}; (2) cosmic gamma-ray emission (diffuse, point-source, and isotropic components) estimated using the publicly available Fermi-LAT models described in Section \ref{sec:gpe}; and (3) the GCE. Using the methods described in the respective sections, the expected numbers of events from (1) and (2) in the 75--300 MeV energy range were estimated to be $N_{\rm 1}$=90.2 and $N_{\rm 2}$=18.3, respectively. Since (1), the BG, is estimated directly from the actual flight data, as described in Section \ref{sec:ds}, its systematic uncertainty is estimated to be less than 1\%. The systematic uncertainty associated with (2) due to the uncertainty in the angular-resolution evaluation described in Section \ref{sec:vela} is estimated to be approximately 8\%. In addition, (2) has uncertainties associated with the Fermi-LAT models. However, since (1) dominates the expected number of events, these systematic uncertainties are negligible compared with the statistical uncertainty in the total expected number of events. Therefore, considering only statistical uncertainties, the expected number of events from (1) and (2) was estimated to be $N_{\rm 1}+N_{\rm 2}$=108.5$\pm$10.4. The number of events detected in the flight data was $N_{\rm obs}=97$, and no statistically significant GCE signal was detected. We determined the upper limit on the number of GCE events based on the observed event count. Specifically, we define $CL_{s+b}$ as the probability of observing $N_{\rm obs}$ or fewer events, assuming a Poisson distribution with an expected number of events of 108.5+$N_{\rm 3}$, where $N_{\rm 3}$ represents the number of GCE events. However, using $CL_{s+b}$ alone may result in an overly stringent upper limit when the observed event count fluctuates below the expectation from (1) and (2). Following the method proposed in \cite{Read}, we therefore define $CL_s=CL_{s+b}/CL_b$, where $CL_b$ is the probability of observing $N_{\rm obs}$ or fewer events, assuming a Poisson distribution with an expected number of events of 108.5. We determine the upper limit by finding the value of $N_{\rm 3}$ for which $CL_s=0.0228$, corresponding to a one-sided 2$\sigma$ confidence level. This gives $N_{\rm 3}$=18.5 events, providing a conservative upper limit on the number of GCE events. Based on this upper limit, we then derive the corresponding upper limit on the GCE flux. The detector response used to convert the number of events into the flux was the one validated in the analysis of gamma-ray emission along the Galactic plane described in Section \ref{sec:gpe}. The resulting conservative upper limit of 1.70$\times$10$^{-7}$ GeV cm$^{-2}$ s$^{-1}$ at the 2$\sigma$ confidence level for the 1$^\circ$-radius ROI centered on the Galactic Center is shown in Figure \ref{fig:GCE}. It should be noted, however, that the systematic uncertainty in the effective-area estimation introduces an approximately 10\% uncertainty in the derived upper limit.

\subsection{Discussion and future prospect}
We compared the obtained upper limit with the spectra predicted for the millisecond-pulsar and dark matter scenarios based on the globular cluster spectra and the DM model, respectively. A very hard spectrum in the sub-GeV energy range, such as that of NGC 6440, is not favored. This result is consistent with the GCE spectrum within 1$^\circ$ of the Galactic Center, estimated from GCE spectra obtained from analyses of Fermi-LAT data using wide ROIs. On the other hand, the spectra of other globular clusters, such as Terzan 5 and NGC 104, and the average spectrum of millisecond pulsars detected by Fermi-LAT, which are somewhat softer than that of NGC 6440, cannot be excluded. Therefore, the present result is consistent with both the millisecond-pulsar and dark matter scenarios for the origin of the GCE.

As shown in Figure \ref{fig:GCE}, the systematic uncertainties in the Fermi-LAT data analysis at several hundred MeV are very large, highlighting the importance of further improving the observational sensitivity of GRAINE toward elucidating the origin of the GCE. Since the upper limit obtained from GRAINE 2023 is predominantly limited by statistical uncertainties, increasing the observation statistics is necessary. The next GRAINE balloon experiment is scheduled to be conducted by the JAXA scientific ballooning group in spring 2028. In GRAINE2028, we plan to launch an instrument of the same design as that used in GRAINE 2023. Combining the GRAINE2028 data with the GRAINE 2023 data presented in this paper will increase the statistics by a factor of two or more. The estimated 2$\sigma$ sensitivity to the GCE based on the increased statistics is also shown in Figure \ref{fig:GCE}. It is derived using twice the 1$\sigma$ statistical uncertainty in the total expected number of events from (1) and (2), combining the present data with those expected from the next experiment. With this sensitivity, we expect to be able to investigate even softer spectra, such as that of Terzan 5 and the average spectrum of millisecond pulsars detected by Fermi-LAT. On the other hand, further improvement in sensitivity is required to clearly distinguish between the millisecond-pulsar and DM scenarios. This requires a larger telescope and a longer observation duration. We are therefore considering a large-scale, long-duration balloon experiment using a NASA balloon flight. Assuming an aperture area of 10 m$^2$, four times larger than that of GRAINE 2023, and an observation duration of one week, seven times longer than that of GRAINE 2023, we estimated the 3$\sigma$ sensitivity to the GCE for this experiment, GRAINE-US, and the resulting sensitivity is also shown in Figure \ref{fig:GCE}. If this observation is realized, GRAINE-US is expected to detect with statistical significance a spectrum in the sub-GeV energy range that is significantly harder than that predicted by the DM scenario, as expected under the millisecond-pulsar scenario. Such a detection is expected to represent significant progress toward elucidating the origin of the GCE. We are currently working on the development efforts required to conduct the GRAINE-US experiment in the early 2030s.

Furthermore, because the spatial distribution of the GCE predicted by the DM scenario, particularly for an NFW profile, is much more centrally concentrated than that of diffuse gamma-ray emission, the GCE-to-diffuse emission ratio is expected to increase further by narrowing the observation region to smaller than the 1$^\circ$ region considered in this paper, enabling a more direct measurement of the GCE flux. Taking advantage of GRAINE's angular resolution in the GeV energy range (0.06$^\circ$ at 2 GeV), we also aim in future experiments to measure the region within 0.1$^\circ$ of the Galactic Center in the 2--3 GeV energy range. The sub-GeV analysis presented in this paper is more sensitive to the millisecond-pulsar scenario because of its harder spectrum, whereas observations within 0.1$^\circ$ of the Galactic Center in the GeV energy range are expected to be more sensitive to the DM scenario because of its stronger central concentration. By pursuing these complementary observations, we expect to approach an understanding of the origin of the GCE.

\section{Conclusion}
The Galactic Center is an important target for understanding the formation and evolution of our Milky Way Galaxy and, more broadly, the diverse population of galaxies in the Universe. In the sub-GeV to GeV high-energy gamma-ray band, an unidentified gamma-ray emission known as the Galactic Center Excess (GCE) has been observed. The GCE has attracted considerable interest because this emission can be explained by gamma rays produced by dark matter annihilation. However, it can also be explained by a population of millisecond pulsars, and its origin remains unresolved. This is primarily due to the large systematic uncertainties associated with contamination from diffuse Galactic gamma-ray emission in current observational data. 

The GRAINE experiment conducts high-resolution observations of cosmic gamma rays using a balloon-borne gamma-ray telescope with emulsion films as its detector. GRAINE achieves an angular resolution of 1$^\circ$ at 100 MeV and 0.1$^\circ$ at 1 GeV, compared with 5$^\circ$ and 0.8$^\circ$, respectively, for Fermi-LAT. By taking advantage of its high angular resolution, GRAINE can conduct observations focused on a region close to the Galactic Center, thereby reducing contamination from diffuse gamma-ray emission in observations of the GCE. The GRAINE 2023 balloon experiment was conducted in Australia in 2023, marking the first observation of the region around the Galactic Center with GRAINE. In this experiment, the aperture area of the telescope was increased by a factor of 6.6 compared with that of the balloon experiment conducted in 2018. To verify the observation performance of GRAINE 2023, we analyzed the Vela pulsar and confirmed that the angular resolution and detector response were as expected. We then proceeded with the analysis of the region around the Galactic Center. Using data from a wide region, we searched for gamma-ray emission along the Galactic plane and detected it with a significance of 6.2$\sigma$. This demonstrated, for the first time with GRAINE, the detection of cosmic gamma-ray emission from the Galactic plane. Furthermore, the observed distribution of gamma-ray emission along the Galactic plane was found to be consistent with the distribution expected from the angular resolution and detector response evaluated in the Vela pulsar analysis. This further demonstrates the observation performance of the GRAINE 2023 flight data for the region around the Galactic Center.

To discuss the results of the GCE search using the flight data, we first calculated the predicted GCE spectra within 1$^\circ$ of the Galactic Center. Assuming dark matter as the origin of the GCE (the DM model), we calculated the DM model spectrum for a dark matter mass of 50 GeV, a velocity-averaged annihilation cross section of 1.76$\times$10$^{-26}$ cm$^3$ s$^{-1}$, and a spatial distribution following an Navarro-Frenk-White (NFW) profile. Assuming millisecond pulsars as the origin of the GCE (the MSP model), we calculated the MSP model spectra using the average spectrum of millisecond pulsars observed by Fermi-LAT and the observational results for several globular clusters considered to host millisecond-pulsar emission. In this analysis, we used only the spectral shapes of the MSP models and scaled their normalizations to match the DM model. The absolute normalization of the MSP emission depends on the number of millisecond pulsars in the region around the Galactic Center. Therefore, the spatial distribution of millisecond pulsars in the Galactic Center region requires further investigation. In addition, we estimated the GCE spectrum within 1$^\circ$ of the Galactic Center based on GCE spectra obtained from Fermi-LAT analyses using wide ROIs.

We searched for the GCE using the GRAINE 2023 flight data. In particular, we focused on the spectral shape in the sub-GeV energy range, which is one of the key aspects for elucidating the origin of the GCE. We compared the number of events detected within 1$^\circ$ of the Galactic Center in the energy range of 75--300 MeV with the estimated contributions from background events dominated by atmospheric gamma rays and from diffuse, point-source, and isotropic gamma-ray emission based on the publicly available Fermi-LAT models. Although no statistically significant GCE was found in the search, we derived a conservative upper limit on the GCE flux of 1.70$\times$10$^{-7}$ GeV cm$^{-2}$ s$^{-1}$ at the 2$\sigma$ confidence level for the 1$^\circ$-radius ROI centered on the Galactic Center based on the analysis results. The obtained upper limit suggests that a very hard GCE spectrum in the sub-GeV energy range, such as that of NGC 6440, is not favored. This result is also consistent with the GCE spectrum estimated from analyses of Fermi-LAT data using wide ROIs. With the current statistics, the results are consistent with both the DM and MSP models for the origin of the GCE. Although GRAINE 2023 did not detect the GCE with statistical significance, the ability to search for the GCE in a small region around the Galactic Center, enabled by GRAINE's high angular resolution, is a unique feature of GRAINE that is difficult to achieve with other existing experiments. We also estimated the expected observational sensitivity of the next GRAINE balloon experiment scheduled for 2028 and of a future large-scale experiment. By pursuing observations with GRAINE in the sub-GeV to GeV energy range while reducing contamination from diffuse gamma-ray emission, we aim to elucidate the origin of the GCE.

\section*{Acknowledgement}
We thank C. Ikeda and the staff of the Scientific Ballooning Research and Operation Group of ISAS/JAXA for providing the scientific balloon (DAIKIKYU) flight opportunity and GPS data, and for their support during the balloon campaign. This work was supported by JSPS KAKENHI Grant Numbers 17H06132, 18H05210, 23H00116, 21H04472, 20H01915, 24K00661, 22K20382, 22KJ2237, 24KJ1273, 25K17437 and 26KJ0149. The Nohmura Foundation for Membrane Structure's Technology; the DAIKO FOUNDATION; the FOUNDATION OF PUBLIC INTEREST OF TATEMATSU; the Nagoya University KMI Overseas Dispatch Program for Young Researchers; and the Nagoya University IMaSS Joint Research Program.

The authors used ChatGPT (OpenAI) for language editing to improve the readability and clarity of the manuscript. All AI-assisted text was reviewed and, where necessary, edited by the authors, who take full responsibility for the content and accuracy of the manuscript.


\begin{thebibliography}{99}
\bibitem{fermi}S. Abdollahi et al., 2009, ApJ, 697, 1071
\bibitem{fermibubble}M. Su, T. R. Slatyer and D. P. Finkbeiner, 2010, ApJ, 724, 1044
\bibitem{GCE1}L. Goodenough and D. Hooper, 2009, arXiv e-prints, arXiv:0910.2998
\bibitem{GCE2}V. Vitale and A. Morselli, 2009, arXiv e-prints, arXiv:0912.3828
\bibitem{DAMPE}F. Alemanno, et al., 2026, ApJS, 284, 22
\bibitem{GCEDM1}D. Hooper and L. Goodenough, 2011, Phys. Lett. B, 697, 412
\bibitem{GCEDM2}D. Hooper and T. Linden, 2011, PhRvD, 84, 123005
\bibitem{GCEDM3}K. N. Abazajian and M. Kaplinghat, 2012, PhRvD, 86, 083511
\bibitem{GCEDM4}D. Hooper and T. R. Slatyer, 2013, Phys. Dark Univ., 2, 118
\bibitem{GCEDM5}T. Daylan, D. P. Finkbeiner, D. Hooper, et al., 2016, Phys. Dark Univ., 12, 1
\bibitem{GCEMSP1}K. N. Abazajian, 2011, JCAP, 2011, 010
\bibitem{GCEMSP2}Q. Yuan and B. Zhang, 2014, JHEAp, 3-4, 1-8
\bibitem{GCEMSP3}R. Bartels, S. Krishnamurthy and C. Weniger, 2016, PRL 116, 051102
\bibitem{GCEMSP4}R. Bartels, E. Storm, C. Weniger and F. Calore, 2018, Nature Astronomy, 2, 819
\bibitem{GCEMSP5}A. Gautam, R. M. Crocker, L. Ferrario, et al., 2022, Nature Astronomy, 6, 703
\bibitem{GRAINE}S. Takahashi, et al., 2018, Adv. Space Res. 62, 2945
\bibitem{GRAINE_polar}K. Ozaki, et al., 2016, Nucl. Instrum. Meth. A 833, 165
\bibitem{GRAINE 2018_1}S. Takahashi, et al., 2023, ApJ. 960, 47
\bibitem{GRAINE 2018_2}Y. Nakamura, et al., 2021, Prog. Theor. Exp. Phys. 2021, 123H02
\bibitem{GRAINE_shifter1}S. Takahashi, et al., 2010, Nucl. Instrum. Meth. A, 620, 192
\bibitem{GRAINE_shifter2}M. Oda, et al., 2022, Prog. Theor. Exp. Phys. 2021, 113H03
\bibitem{GRAINE_gondola}H. Rokujo et al., 2026, arXiv e-prints, arXiv:2608.29050
\bibitem{GRAINE_gamma}H. Rokujo, et al., 2018, Prog. Theor. Exp. Phys., 2018, 063H01
\bibitem{geant}S. Agostinelli. et al., 2003, Nucl. Instrum. Meth. A, 506, 250
\bibitem{GRAINE_precise}Y. Nakamura, et al., 2025, Astroparticle Phys., 165, 103055
\bibitem{fermiVela}A. A. Abdo et al., 2010, ApJ, 713, 154
\bibitem{fermi_iem}Fermi-LAT Collaboration, Galactic Interstellar Emission Model for the 4FGL Catalog Analysis, 2019
\bibitem{fermi_ps}J. Ballet, et al., 2022, ApJS, 260, 53
\bibitem{wilks}S. S. Wilks, 1938, Ann. Math. Statist. 9, 1, 60
\bibitem{TS}J. R. Mattox, 1996, ApJ, 461, 396
\bibitem{GCE_calore}F. Calore, et al., 2015, Phys. Rev. D 91, 063003
\bibitem{PPPC4}M. Cirelli, et al., 2011, JCAP, 1103, 051
\bibitem{GCE_fermi1}F. Calore, I. Cholis and C. Weniger, 2015, JCAP, 03, 038
\bibitem{GClist}D. Song, et al., 2021, MNRAS, 507, 5161
\bibitem{AllMSP}I. Cholis, D. Hooper, T. Linden, 2014, arXiv:1407.5583
\bibitem{GCE_fermi2}C. Gordon and O. Macias, 2013, Phys. Rev. D, 88, 083521
\bibitem{GCE_fermi3}M. Ackermann, et al., 2017, ApJ, 840, 43
\bibitem{Read} A L Read, 2002, J. Phys. G: Nucl. Part. Phys., 28, 2693

\end{thebibliography}
\end{document}